\documentclass[a4paper,11pt]{article}
\usepackage{pos}

\title{\texttt{FORM} 5.0 and further developments}

\author[a]{Ana Costa Pereira}
\author*[a]{Joshua Davies}

\affiliation[a]{Department of Mathematical Sciences, University of Liverpool\\
Liverpool, L69 3BX, UK}

\emailAdd{Ana.Costa-Pereira@liverpool.ac.uk}
\emailAdd{j.o.davies@liverpool.ac.uk}

\abstract{
In these proceedings we review the features of the recently-released version 5.0 of the \texttt{FORM} computer algebra system, which is one of the core software packages for the large-scale symbolic manipulations required by calculations in theoretical high-energy physics. We also discuss developments since the release of version 5.0, focussing in particular on recent performance improvements.
}

\FullConference{23rd International Workshop on Advanced Computing and Analysis Techniques in Physics Research (ACAT2025)\\
8–12 September 2025\\
Hamburg, Germany\\}

\begin{document}
\maketitle

\section{Introduction}

The \texttt{FORM} computer algebra system has been a key software package for multi-loop and multi-scale computations in perturbative quantum field theory for several decades \cite{Vermaseren:2000nd,Tentyukov:2007mu,Kuipers:2012rf,Kuipers:2013pba,Ruijl:2017dtg,Ueda:2020wqk}. Its design allows it to efficiently process expressions which are larger than computer memory, enabling computations which are not possible with other computer algebra systems. Although memory sizes have grown enormously over the years, so too has the scale of cutting-edge theory computations; thus \texttt{FORM} has had a large impact on the success of particle physics endeavours. Since version 4.0 (2012) \cite{Kuipers:2012rf}, \texttt{FORM} has been released as free software (GPLv3), meaning that large-scale computations can be run across computing clusters without restrictions, for example due to license availability at user research institutions.

\texttt{FORM} 5.0 \cite{Davies:2026cci} has been released recently. In these proceedings we summarize the features of the new release, however we refer the reader to Ref.~\cite{Davies:2026cci} for a much more detailed description. We also discuss recent performance improvements which are available in version 5.0.1, which can be obtained from \texttt{FORM}'s development repository, hosted at:

\begin{center}
	\url{https://github.com/form-dev/form}
\end{center}

\section{New features}

\texttt{FORM} 5.0 contains a number of features which provide new capabilities, described below, or provide usability improvements of existing features including more robust \texttt{Fortran}- and \texttt{Mathematica}-formatted output, additional warning messages, easier-to-read sorting statistics, and more flexibility regarding the placement of \texttt{ModuleOption} statements for dollar variables. The new \texttt{\#continuedo} statement allows for more complex pre-processor loop construction. A \texttt{TableBase} may now be opened in \texttt{readonly} mode, which is useful when providing access to files on a shared network filesystem. Temporary sort files are now compressed with \texttt{zstd} \cite{rfc8878} if possible, providing a performance improvement particularly when using slower storage devices. The new compilation option \texttt{--enable-backtrace} provides additional information during crashes, to be used for bug reports. For more information, we refer the reader to Ref.~\cite{Davies:2026cci}.

\subsection{Built-in graph generator}
\texttt{FORM} 5.0 features a built-in Feynman graph generator, based on a \texttt{C++} port of the generator from the \texttt{GRACE} system \cite{Kaneko:1994fd}. \texttt{FORM}-script syntax has been introduced to define a Lagrangian; between \texttt{Model} and \texttt{EndModel} statements, one may define \texttt{Particle} and \texttt{Vertex} types which define the content, interactions and coupling constants of the model, for example:
\vspace{-2mm}\begin{verbatim}
    Model PHI3;
        Particle phi, +1;
        Vertex phi,phi,phi: g;
    EndModel;
\end{verbatim}
\vspace{-2mm}
Once a model has been defined, graphs may be generated using the \texttt{diagrams\_} function, which produces output as a sequence of \texttt{FORM} terms which can be further manipulated by the script, using the standard \texttt{FORM} features. The graphs which are generated may be filtered using a variety of keywords, such as \texttt{OnePI\_} or \texttt{NoTadpole\_} etc, which are fully described in Ref.~\cite{Davies:2026cci}. Including a generator directly in \texttt{FORM} removes the need to generate graphs in external software and convert the output to compatible syntax, and provides a framework for the user to extend the generator's functionality; its \texttt{C++} code is covered by the same GPLv3 license as the rest of \texttt{FORM}'s codebase.

\subsection{Floating-point coefficients}
The new pre-processor command \texttt{\#StartFloat Nd}\, enables the new arbitrary-precision floating-point coefficient mode, with a precision of \texttt{N} digits. By using the GMP~\cite{Granlund12} and MPFR~\cite{mpfr} libraries, this provides the ability to represent term coefficients as floating-point numbers as well as the numerical evaluation of a variety of built-in functions, including \texttt{sqrt\_}, \texttt{ln\_}, trigonometric functions, multiple zeta values (\texttt{mzv\_}) and Euler sums (\texttt{euler\_}). Rational numbers may be converted to floating-point values with the \texttt{ToFloat} statement, and floating-point values back to rational numbers with \texttt{ToRational} using an algorithm based on continued fractions.

\subsection{\texttt{FLINT} interface}
\label{sec:flint}
The \texttt{FLINT} library \cite{flint} provides high-performance routines for polynomial arithmetic. \texttt{FORM} 5.0 features an interface to \texttt{FLINT}; \texttt{FORM}'s internal term representation is converted into \texttt{FLINT} polynomial types, the required computation is performed by the library, and the result is converted back. The use of \texttt{FLINT} is enabled by default, does not require any changes to user scripts, and provides a large performance improvement for computations which make use of
\texttt{div\_},
\texttt{rem\_},
\texttt{mul\_},
\texttt{gcd\_}, and
\texttt{inverse\_}, as well as
\texttt{PolyRatFun},
\texttt{FactArg} and
\texttt{FactDollar}.
Benchmarks involving univariate polynomials see run-time improvements from 1.2x to 2x, and heavier multivariate benchmarks up to 50x or more, depending on the number of variables. Computations involving many variables are much more common than they were historically; this interface makes \texttt{FORM} much faster for modern particle physics, ``\texttt{MaxTermSize}''-related issues notwithstanding\footnote{Version 5.0.1 reduces \texttt{FORM}'s large memory allocation size for larger values of \texttt{MaxTermSize} to some extent. Further improvements in this direction are ongoing.}.

\section{Performance improvements}
Recently, we have started collecting a series of \texttt{FORM} codes which serve as performance benchmarks. One of the main aims has been for benchmark codes to represent realistic \texttt{FORM} workloads, thus many of the tests are based on public \texttt{FORM} packages including
\texttt{color} \cite{vanRitbergen:1998pn},
\texttt{fmft} \cite{Pikelner:2017tgv},
\texttt{forcer} \cite{Ruijl:2017cxj},
\texttt{hyperform} \cite{Kardos:2025klp},
\texttt{mincer} \cite{Larin:1991fz}
and code from the Multiple Zeta Value data mine \cite{Blumlein:2009cf}. The repository containing the benchmarking suite is hosted at:

\begin{center}
	\url{https://github.com/form-dev/form-bench}
\end{center}

The benchmarking suite has enabled systematic performance profiling and improvement efforts, which we summarize here. Some improvements apply to features introduced in \texttt{FORM} 5.0 (in particular, further improvements to the \texttt{FLINT} interface), while others are changes to long-standing, widely-used code. The widespread use of \texttt{FORM} for demanding computations means that these improvements enable significant savings in the CPU time required by future computations. Improvements come from the changes described below. The performance impact of each change depends strongly on which benchmark is considered; each stresses different parts of the \texttt{FORM} system. The impact also depends on the system running the benchmarks\footnote{Here, an AMD Ryzen 9 7900X with 32GB DDR5-6000. The CPU is configured for reliable benchmarking by limiting its maximum frequency to 5.1GHz. Benchmarks are run on an idle system at high priority, with temporary files stored in RAM to remove disk performance and page caching effects.}.

\vspace{0.5em}
\noindent\textbf{Improved data-copying loops:} profiling reveals several running-time hotspots which copy data between memory buffers, for example when replacing sub-expression pointers with their expression data and when transferring term data between threads in \texttt{TFORM}. Optimizations involve hoisting conditionals out of hot loops, resulting in a 0-8\% improvement.

\vspace{0.5em}
\noindent\textbf{Scratch area for term normalization:} the frequently-called \texttt{Normalize} function is responsible for putting terms into their canonical form. Historically, in an effort to maintain good performance, it has used stack-allocated memory for its temporary scratch area used to construct the canonical term result. However, large stack allocations in frequently called functions can suffer due to security features implemented in modern compilers; here, \textit{stack clash protection}. Moving the scratch area to a dynamically-allocated array stored in the global data produces a 0-33\% performance improvement.

\vspace{0.5em}
\noindent\textbf{Improved parallel sorting:} in \texttt{TFORM}, the ``sortbot'' mechanism is intended to parallelise the final merge of the sorted streams of terms from each worker thread. The worker and sortbot threads transfer data by means of ring-buffer data structures, accessed in parallel by data-writing and data-reading threads. Over the years, these buffers have grown in default size, resulting in inefficient, or a lack of, parallelization for small- to medium-scale computations. By having data-writing threads detect dynamically if a data-reading thread is stalled, waiting for data, they are able to yield the current ring-buffer slot to the data-reading thread as soon as possible, rather than continuing to completely fill the current slot. This results in a 0-35\% improvement for \texttt{TFORM}.

\vspace{0.5em}
\noindent\textbf{Removed locking for dollar variables:} previously, the use of dollar variables in \texttt{TFORM} with \texttt{ModuleOption minimum/maximum} modes involved mutex locks on read and write accesses. Making these dollar variable types thread-local and collecting their global minimum/maximum values at the end of the module results in a 2-25\% improvement in some benchmarks, particularly when running \texttt{TFORM} with larger numbers of workers.

\vspace{0.5em}
\noindent\textbf{\texttt{FLINT} interface and polynomial arithmetic:} in addition to the performance improvements due to the use of \texttt{FLINT} discussed above in Section~\ref{sec:flint}, further improvements are made by using a lexicographic variable ordering such that input (output) term sequences to (from) \texttt{FLINT} functions can be used directly without intermediate sorts from (to) \texttt{FORM}'s canonical term ordering. Benchmarks which make heavy use of multivariate polynomial arithmetic see an improvement of 5-31\%.

\vspace{0.5em}
For the current set of tests contained in the benchmarking suite, the combined effect of the improvements w.r.t.~version 5.0 described above is as follows, when run with \texttt{TFORM} and 12 worker threads, on the system described in the footnote. Running times and uncertainties are measured using \texttt{hyperfine} \cite{Peter_hyperfine_2023}. These performance improvements are contained in the version 5.0.1 patch release, available from \texttt{FORM}'s GitHub repository.

\vspace{1em}
\noindent
	\begin{minipage}{0.329\textwidth}
	\begin{flushright}
	\begin{tabular}{|l|l|}
	\hline 
	\textbf{Benchmark} & \textbf{Speedup} \\ 
	\hline 
	\texttt{chromatic} & 1.24 ± 0.01 \\ 
	\texttt{color} & 1.24 ± 0.02 \\ 
	\texttt{fmft} & 1.30 ± 0.01 \\ 
	\texttt{forcer} & 1.32 ± 0.01 \\ 
	\texttt{forcer-exp} & 1.37 ± 0.01 \\ 
	\texttt{hyperform} & 1.21 ± 0.01 \\ 
	\texttt{ibp-box2l} & 1.46 ± 0.02 \\ 
	\hline 
	\end{tabular}
	\end{flushright}
	\end{minipage}
	\begin{minipage}{0.329\textwidth}
	\begin{center}
	\begin{tabular}{|l|l|}
	\hline 
	\textbf{Benchmark} & \textbf{Speedup} \\ 
	\hline
	\texttt{ibp-npbox2l} & 1.36 ± 0.01 \\ 
	\texttt{ibp-mbox1l} & 1.64 ± 0.05 \\
	\texttt{ibp-pent1l} & 1.71 ± 0.06 \\
	\texttt{ibp-tri3l} & 1.03 ± 0.00 \\
	\texttt{ibp-nptri3l} & 1.05 ± 0.01 \\
	\texttt{minceex} & 1.23 ± 0.01 \\ 
	\texttt{mincer} & 1.35 ± 0.01 \\ 
	\hline
	\end{tabular}
	\end{center}
	\end{minipage}
	\begin{minipage}{0.329\textwidth}
	\begin{flushleft}
	\begin{tabular}{|l|l|}
	\hline 
	\textbf{Benchmark} & \textbf{Speedup} \\ 
	\hline 
	\texttt{mzv-dm} & 1.28 ± 0.02 \\ 
	\texttt{sort-disk} & 1.35 ± 0.04 \\ 
	\texttt{sort-large} & 1.75 ± 0.02 \\ 
	\texttt{sort-small} & 1.81 ± 0.04 \\ 
	\texttt{trace} & 1.35 ± 0.01 \\ 
	&\\
	&\\
	\hline 
	\end{tabular}
	\end{flushleft}
	\end{minipage}

\section{Conclusions and outlook}
\texttt{FORM} 5.0 has provided a variety of new features, fixes of long-standing bugs, and performance improvements, enhancing its utility for modern high-precision computations in theoretical physics. Development has seen a large increase in momentum in recent years, driven primarily by the series of annual development workshops which started in 2023. These have provided opportunities to bring developers and users together to discuss the future of \texttt{FORM}, by providing technical information about its codebase and how it may be modified, and collecting information on the features desired by the user community; these include facilities for the reconstruction of expressions from finite-field samples, variable namespacing to enable the development of more robust packages, further-improved rational polynomial systems, a $p$-adic coefficient mode, improved CI testing and reliability, easy-to-use software and package distribution via \texttt{PyPI} and an updated reference manual and tutorial.
We strongly encourage readers with an interest in \texttt{FORM}'s future to attend these workshops and participate actively in development efforts, ensuring its long-term availability as the high-performance computer algebra package that theoretical particle physics requires.

\bibliographystyle{JHEP}
\bibliography{acat-2025-form.bib}
\end{document}